\documentclass[a4paper]{article}

\usepackage{amsmath}          
\usepackage{amstext}          
\usepackage{amssymb}          
\usepackage{amsfonts}         
\usepackage{hyperref}

\renewcommand{\P}{\operatorname{\mathbb{P}}}
\newcommand{\Ai}{\operatorname{Ai}}
\newcommand{\dd}{\mathrm{d}}
\newcommand{\ee}{\mathrm{e}}

\title{Comment on the paper ``Exact decay of the persistence probability in the Airy$_1$ process'' by Ferrari and Liu}
\author{Sylvain Prolhac}
\date{Laboratoire de Physique Th\'eorique, UPS, Universit\'e de Toulouse, France\\\vspace{5mm}December 11, 2024}

\begin{document}
\maketitle

\begin{abstract}
We point out that the non-trivial function obtained by Ferrari and Liu for the persistence probability of the Airy$_1$ process has a strikingly similar form as a large deviation function found earlier by the author for current fluctuations of the totally asymmetric exclusion process with periodic boundaries conditioned on flat initial and final states. A proposed explanation for this observation, which relates similar yet clearly distinct quantities, is that both results pertain to conditioning on the same kind of rare events where a current larger than typical is maintained throughout the system.
\end{abstract}

\vspace{5mm}

The paper \cite{FL2024} mentioned in the title studies the Airy$_1$ process $\mathcal{A}_{1}(x)$, which describes universal spatial correlations of one-dimensional growing interfaces at the KPZ fixed point, when the infinitely long interface is initially completely flat. More precisely, the persistence probability for the Airy$_1$ process is proved in \cite{FL2024} to vanish as $\P(\mathcal{A}_{1}(x)\leq u,\,x\in[0,\ell])\simeq A\,\ee^{-2\ell\varphi(2u)}$ at large $\ell$, with $A$ a constant and $\varphi$ a non-trivial function known explicitly, see below.

Interestingly, the very same function $\varphi$ had appeared earlier \cite{P2015} for the one-dimensional totally asymmetric simple exclusion process (TASEP), a Markov process where driven particles on a lattice move from any site $i$ to the next one $i+1$ with unit rate, and which converges to the KPZ fixed point at large scales. More precisely, it was found in \cite{P2015} that $\varphi$ describes large deviations of current fluctuations for TASEP with periodic boundaries conditioned on specific initial and final states corresponding to a flat interface in the KPZ language.

From \cite{BFPS2007}, the result of \cite{FL2024} mentioned above about persistence for the Airy$_1$ process can be rewritten in terms of TASEP \emph{on the infinite line} $\mathbb{Z}$ prepared at time $t=0$ in the \emph{flat state} (in the growing interface language) where even sites are occupied and odd sites are empty, as
\begin{equation}
	\label{P Qi}
	\P\Big(-\frac{4}{t^{1/3}}\big(Q_{L}^{\rm min}(t)-\tfrac{t}{4}\big)\leq u\Big)\simeq A\,\ee^{-\frac{L}{t^{2/3}}\,\varphi(u)}
\end{equation}
when $1\ll t\ll L^{3/2}$, with $Q_{L}^{\rm min}(t)=\min\{Q_{i}(t),\,i=1,\ldots,L\}$ and $Q_{i}(t)$ the total number of particles that have moved from site $i$ to site $i+1$ between time $0$ and time $t$.

On the other hand, \cite{P2015} considers TASEP with $L$ sites and \emph{periodic boundaries}, for an evolution \emph{conditioned on beginning and ending in the flat state} (defined as above). Then, in terms of the mean current $\overline{Q}(t)=\frac{1}{L}\sum_{i=1}^{L}Q_{i}(t)$ with $Q_{i}(t)$ as above, one has
\begin{equation}
	\label{P Qbar}
	\P\Big(-\frac{4}{t^{1/3}}\big(\overline{Q}(t)-\tfrac{t}{4}\big)\in[u,u+\dd u]\Big)\simeq \tfrac{B\sqrt{L}}{t^{1/3}}\,\ee^{-\frac{L}{t^{2/3}}\,(\varphi(u)+\frac{3u}{2}+C)}\,\dd u\;,
\end{equation}
in the regime $1\ll t\ll L^{3/2}$, with $B$ and $C$ known constants.

The striking resemblance of the leading contribution in (\ref{P Qi}) and (\ref{P Qbar}) can be motivated physically by noting that both cases evaluate the probability of rare events where the current is maintained larger than typical throughout a large region of $L$ sites (which is arbitrary in the former case, and equal to the system size in the latter case) : this is true in (\ref{P Qi}) since large enough negative fluctuations of the $Q_{i}(t)$ are suppressed at any site $i$ within the region considered, while in (\ref{P Qbar}) the system being close to the flat state at the end of the evolution ensures that particles are not blocked by their neighbours then and leads to a current higher than for the free evolution. Additionally, the regime $t\ll L^{3/2}$ ensures that the correlation length, which grows as $t^{2/3}$, stays much smaller that the system size in (\ref{P Qbar}), so that no difference is expected between TASEP on $\mathbb{Z}$ and TASEP on $L$ sites with periodic boundaries here.

The leading contributions in (\ref{P Qi}) and (\ref{P Qbar}) are related to the function
\begin{equation}
	\psi(u)=-\sum_{n=1}^{\infty}\frac{\Ai'(u\,n^{2/3})}{n^{5/3}}\;,
\end{equation}
with $\Ai$ the Airy function, which is analytic except for discontinuities of its derivative at $u\in-(3\pi\mathbb{N})^{2/3}$. Analytic continuation, which can be performed explicitly \cite{FL2024,P2015}, leads in particular to $\psi_{+}(u)$ and $\psi_{-}(u)$ analytic for $u\in\mathbb{R}$ and coinciding with $\psi(u)$ respectively for $u>0$ and $-(3\pi)^{2/3}<u<0$, and related by $\psi_{-}(u)=\psi_{+}(u)+\frac{3u}{2}$. Then, the leading contributions in (\ref{P Qi}) and (\ref{P Qbar}) are respectively
\begin{equation}
	\varphi(u)=\psi_{+}(u)\;,
	\quad\text{and}\quad
	\varphi(u)+\frac{3u}{2}+C=\psi_{-}(u)-\psi_{-}(u_{*})
\end{equation}
with $u_{*}\approx-2.050$ the location of the unique minimum of $\psi_{-}$. In particular, (\ref{P Qbar}) indicates that the random variable $\chi(t)=-\frac{4}{t^{1/3}}\big(\overline{Q}(t)-\tfrac{t}{4}\big)$ is equal to $u_{*}$ almost surely, with Gaussian typical fluctuations of standard deviation $\sim t^{1/3}/\sqrt{L}\ll1$. The simple shift by which the leading contributions in (\ref{P Qi}) and (\ref{P Qbar}) differ is then reminiscent of an exponential tilting, corresponding to conditioning $\chi(t)$ on its mean value $u_{*}$.\\

\noindent
\textbf{Acknowledgement} : it is a pleasure to thank P.~Ferrari for exchanges last February about a possible relation between the results in \cite{FL2024} and \cite{P2015}.

\end{document}